\documentclass[useAMS,usenatbib]{mn2e}
\def\farcs{\hbox{$.\!\!^{\prime\prime}$}}
\def\kms      {\ifmmode {\rm km\,s}^{-1} \else km\,s$^{-1}$\fi}

\def\ltsim{\ifmmode\stackrel{<}{_{\sim}}\else$\stackrel{<}{_{\sim}}$\fi}
\def\gtsim{\ifmmode\stackrel{>}{_{\sim}}\else$\stackrel{>}{_{\sim}}$\fi}
\begin{document}

\title[HI Absorption in 3C\,293] {High resolution imaging of the radio 
continuum and neutral gas in the inner kiloparsec of the radio galaxy 
3C\,293} \author[Beswick et al.,]{R. J. Beswick,$^{1}$\thanks{rbeswick@jb.man.ac.uk} A. B. Peck,$^{2}$ G. B. Taylor$^{3}$ and G. Giovannini$^{4}$ \\1. The University of
Manchester, Jodrell~Bank Observatory, Macclesfield, Cheshire SK11~9DL,
UK\\2. Harvard-Smithsonian Center for Astrophysics, SAO/SMA Project, P.O. 
Box 824, Hilo, HI 96721\\3. National Radio Astronomy Observatory, P.O. Box 
O, Socorro, NM 87801\\4. Istituto di Radioastronomia del CNR, via Gobetti 101, 40129 Bologna, Italy}

\date{Accepted 29/3/04 }
       
\pagerange{\pageref{firstpage}--\pageref{lastpage}} \pubyear{2004}

\maketitle
\label{firstpage}
\begin{abstract}

Using a combination of observations involving the VLA, MERLIN and global VLBI
networks we have made a detailed study of the radio continuum and the neutral
hydrogen (H{\sc i}) kinematics and distribution within the central kiloparsec of the
radio galaxy 3C\,293. These observations trace the complex jet structure and identify
the position of the steeply inverted radio core at 1.3\,GHz.

Strong H{\sc i} absorption is detected against the majority of the inner kiloparsec
of 3C\,293. This absorption is separated into two dynamically different and spatially
resolved systems. Against the eastern part of the inner radio jet narrow H{\sc i}
absorption is detected and shown to have higher optical depths in areas co-spatial
with a central dust lane. Additionally, this narrow line is shown to follow a
velocity gradient of $\sim$50\,km\,s$^{-1}$\,arcsec$^{-1}$, consistent with the
velocity gradient observed in optical spectroscopy of ionised gas. We conclude that
the narrow H{\sc i} absorption, dust and ionised gas are physically associated and
situated several kiloparsecs from the centre of the host galaxy. Against the western
jet emission and core component, broad and complex H{\sc i} absorption is detected.
This broad and complex absorption structure is discussed in terms of two possible
interpretations for the gas kinematics observed. We explore the possibility that
these broad, double absorption spectra are the result of two gas layers at different
velocities and distances along these lines of sight. A second plausible explanation
for this absorbing structure is that the H{\sc i} is situated in rotation about the
core of this radio galaxy with some velocity dispersion resulting from in-fall and
outflow of gas from the core region. If the latter explanation were correct, then the
mass enclosed by the rotating disk would be at least 1.7$\times$10$^9$ solar masses
within a radius of 400\,pc.

\end{abstract} 


\section{Introduction}

Nuclear activity in galaxies manifests itself in a variety of forms from
nearby low luminosity active galactic nuclei (AGN), such as Seyferts and
LINERS, to powerful distant quasars and radio galaxies.  In these sources,
the nuclear activity is responsible for radiation detected across the
entire electromagnetic spectrum.  In the radio loud active galaxies, such
as quasars and radio galaxies, the radio emission demonstrates the
influence of the AGN, for example via the formation of powerful jets
(Fanaroff \& Riley 1974).  Additionally ample evidence is also available
from other wavelength ranges ({\it e.g} optical) for the interaction of
the nuclear activity with the surrounding galactic interstellar medium
(ISM) such as via the detection of outflows from nuclear regions of some
Seyfert galaxies ({\it e.g.} NGC\,3079, Cecil et al. 2001). The commonly
accepted standard model for nuclear activity asserts that the AGN is
fuelled by the release of gravitational potential energy as galactic
material is accreted onto a central super-massive black hole. As such it
is incumbent upon investigators to study not only the effects of this
nuclear activity ({\it e.g.} jets) but also the physical and kinematic
environment that surrounds the AGN, since this provides a method by which
we can study how gas, dust and stars act as fuel for the activity we
observe as well as how this activity impacts the surrounding ISM.

At the present time most high angular resolution studies of powerful
active galaxies have concentrated upon investigating the consequences of
the activity ({\it e.g.} the synchrotron emission such as radio jets),
rather than the cause of the activity ({\it e.g.} cold neutral and molecular
gas that fuels the nuclear activity). This has primarily been a direct
consequence of observational constraints resulting from the relatively
small collecting areas of the current generation of mm-wavelength aperture
synthesis instruments and the surface brightness sensitivity of decimetre
wavelength interferometers precluding observations of cold thermal gas
emission at angular resolutions of $\ltsim$1\,arcsec.  However using
current radio aperture synthesis techniques it is possible to observe cold
gas, via decimetre transitions such as H{\sc i}, OH and H$_2$CO, in
absorption against the bright background radio continuum of some galaxies,
on sub-arsecond angular scales.  
\begin{figure} 
\begin{center}
\setlength{\unitlength}{1mm}
    \begin{picture}(80,75)
\put(0,0){\includegraphics{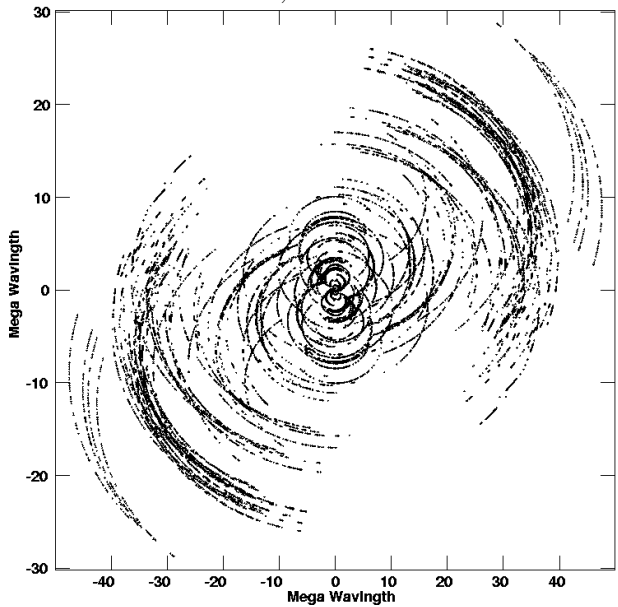}}
\end{picture} 
\caption{The {\it u-v} coverage of the combined VLBI and MERLIN data-sets.} 
\label{fig1a}
\end{center}

\end{figure}
\begin{center}
\begin{table}
\caption[]{Summary of Observations. All three of these observations were
made in spectral line mode with the bandwidth centred at 1.359\,GHz.}
\begin{tabular}{lccc}
\hline
Interferometer&Date&BW&{\it uv} Range\\
&&(MHz)&(k$\lambda$)\\
\hline
VLA-A$+$PT&15/12/2000&3.125&3.1-235\\
MERLIN&08/04/1998&8&11-989\\
VLBI&18/11/1999&8&162-4850\\
\hline
\end{tabular}
 \label{tab1}
\end{table}
\end{center}

3C\,293 is a nearby radio galaxy associated with the peculiar elliptical
galaxy {\sc vv}5-33-12. On scales of several tens of kiloparsecs, the
radio jet structure of 3C\,293 has been well studied by Bridle, Fomalont
\& Cornwell (1981) and van Breugel et al. (1984) and resembles a
moderately large two-sided FR-II radio galaxy.  However, 3C\,293 is
peculiar in that an unusually high proportion of the galaxy's radio power
is emitted from a steep-spectrum extended core component.  This core
region when observed at higher angular resolution, is found to be a
composite of several radio components forming a kiloparsec scale east-west
orientated jet (Bridle et al. 1981; Akujor et al. 1996; Beswick, Pedlar \&
Holloway 2002). At other wavelengths, 3C\,293 and its associated galaxy
display several distinctive characteristics. VV5-33-12 has a closely interacting small companion galaxy situated
$\sim$37\arcsec\,\,($\sim$30\,kpc) toward the south-west (Heckman et al.
1985; Evans et al. 1999) and the central region of the galaxy is
criss-crossed by several filamentary dust lanes aligned in an
approximately north-south direction (van Breugel et al. 1981; Martel el.
1999; Allen et al. 2002). Additionally {\it Hubble Space Telescope} ({\it
HST}) observations have detected an optical/IR jet within the central
kiloparsec (Leahy, Sparks \& Jackson 1999) partially obscured from
previous observations by the nuclear dust lanes. Arcsecond resolution
observations of molecular gas in 3C\,293 by Evans et al. (1999) have
revealed large concentrations of CO(1$\rightarrow$0) detected in both
emission and absorption within the central few kiloparsecs. The CO
emission is primarily distributed in an asymmetric disk rotating about an
unresolved continuum component that Evans et al. conclude is the AGN.

\begin{figure} 
\setlength{\unitlength}{1mm}
\begin{picture}(80,77)
\put(0,0){\includegraphics{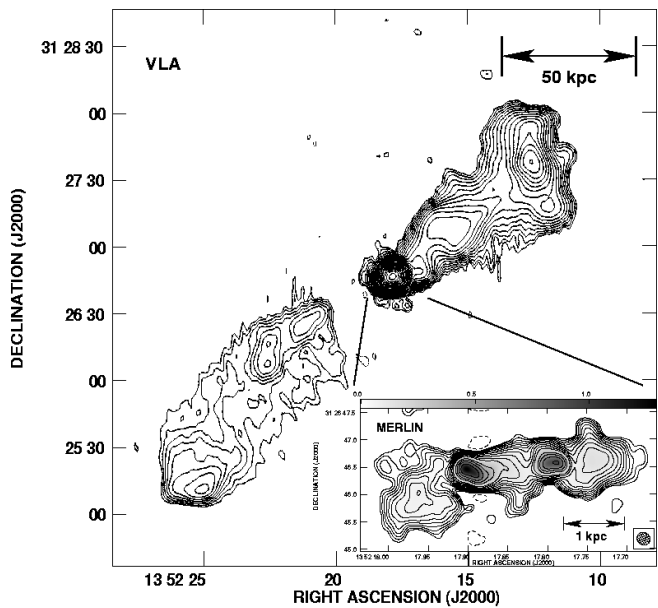}}
\end{picture} 
\caption{VLA B configuration image at 1.35\,GHz of the large scale
structure of the radio galaxy 3C\,293 along with inset MERLIN image of
the central $\sim$3\,kpc. The VLA image has been contoured at $\sqrt2$
times 0.5\,mJy\,beam$^{-1}$ with a peak flux of 3.44\,Jy. The lowest
contour of the inset MERLIN image is 5\,mJy\,beam$^{-1}$ and follows the
same multiplying factors as the VLA image. The peak of the MERLIN image
is 1.29\,Jy\,beam$^{-1}$. The MERLIN radio continuum image can be seen in
more detail in Fig.\,2.}
\label{fig1}
\end{figure}

Broad neutral hydrogen absorption was first detected against 3C\,293 by
Baan \& Haschick (1981) using the Arecibo telescope and has been studied in great detail with ever improving sensitivities and angular
resolutions using a variety of radio interferometers over the last two
decades (Shostak et al. 1983; Haschick \& Baan 1985; Beswick et al. 2002;
Morganti et al. 2003). 3C\,293 has proved to be worthy of these numerous
studies because of its complex and exceptionally broad H{\sc i} absorption
structure which has been observed against the extended nuclear radio
continuum source. In order to fully sample the wide range of physical
scales of the nuclear radio continuum of 3C\,293 from several arcseconds
to angular resolutions of a few tens of mas we have combined global Very
Long Baseline Interferometry (VLBI) observations with previously published
Multi-Element Radio Linked Interferometric Network (MERLIN) data (Beswick
et al. 2002) and Very Large Array (VLA: A configuration including the VLBA
Pie Town antenna) observations to provide a wide range of {\it u-v}
spacings (see Tab.\,1 and Fig.\,1). This combined study allows this
extended radio source to be imaged with high fidelity at a variety of
angular resolutions.  

\begin{figure*} 
\setlength{\unitlength}{1mm}
    \begin{picture}(80,105)
\put(0,0){\includegraphics{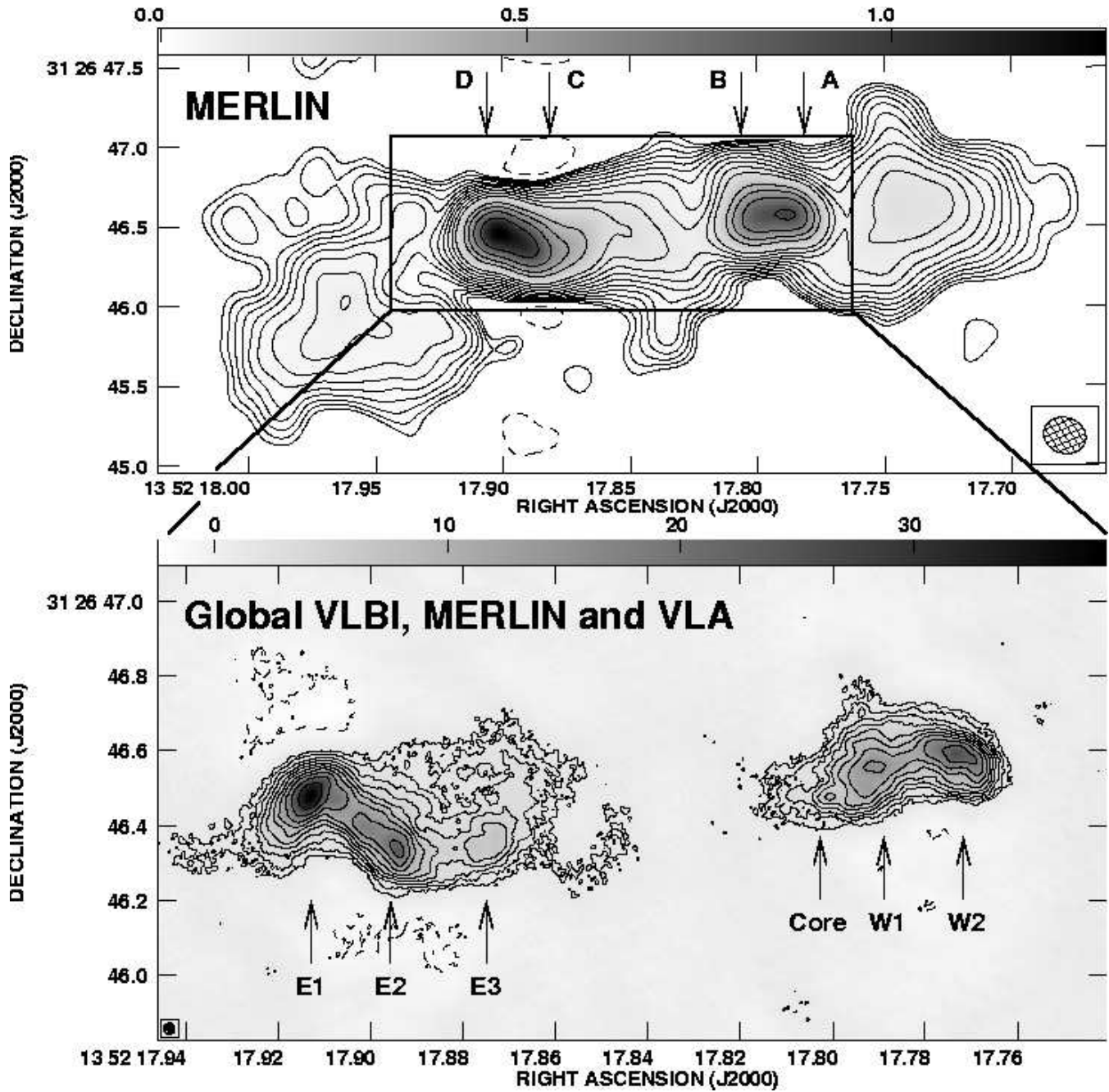}}
\end{picture} 
\vskip 16pc 
\caption{Sub-arcsecond continuum structure of the inner few kiloparsecs
of 3C\,293. The top contour map shows the 1.359 GHz radio continuum
structure observed with MERLIN at a resolution of
0\farcs23$\times$0\farcs20. Contour levels of the MERLIN image are the
same as for Fig.\,1. The lower panel shows the global VLBI, MERLIN and
VLA+PT contoured image of the inner jet of 3C\,293 with angular resolution
of 30\,mas. This map is contoured at multiples of $\sqrt2$ times
1.3\,mJy\,beam$^{-1}$. The peak flux of the 30\,mas image is 38.23\,mJy.
Labels A, B, C and D at the top of the MERLIN image follow the convention
of labeling continuum components in this source used by Bridle et al.
(1981) and Beswick et al. (2002).}
\label{fig2}
\end{figure*}

This paper is split into four additional sections. The first of these will
describe the observations presented and the data processing that has been
applied to them. This will be followed by the presentation of the
observational results and a more detailed discussion of their
implications. The discussion will initially concentrate upon the radio
continuum structure of 3C\,293 from arcmin to milliarcsec scales,
comparing these new observations with previously published data-sets,
followed by a detailed discussion of these new mas resolution H{\sc i}
absorption observations against the central kiloparsec of 3C\,293. The
final section of this paper will outline the key conclusions of this work
and place them in context of other radio galaxies and their environments.

Throughout this paper we assume H$_0=$75\,km\,s$^{-1}$\,Mpc$^{-1}$. At a redshift of z=0.045 this implies a distance for 3C\,293 of 180\,Mpc so an angular size of 1\,mas corresponds to 0.815\,pc. 

\begin{center}
\begin{table*}
\caption[]{Spectral indices between 1.359 and 4.546 GHz\footnotemark[1]
for components in the inner jet. Flux densities for both frequencies have
been obtained from matched angular resolution (50 mas) images.  Positions
of components have been derived from the 1.359 GHz image presented
pictorially in Fig.\,2, bracketed labels equivalent to the position
labels of the spectra in Fig.\,3 \& 4 and Tab.\,3. The flux densities
quoted are for the peaks for each radio continuum component at both
frequencies in addition to the integrated flux density of each component.
Spectral indices have been calculated using
S$_\nu\propto\nu^{-\alpha}$.}
\begin{tabular}{l||c|c|c|c|c|c|c}
\hline
Component&RA (J2000) &Dec (J2000)&S$_{\rm 1.359GHz}$&S$_{\rm 1.359GHz}$&S$_{\rm 4.546GHz}$&S$_{\rm 4.546GHz}$&Spectral\\
&&&Peak&Total&Peak&Total&Index\\
&13$^{\rm h}$\,52$^{\rm m}$&31\degr\,26\arcmin&(mJy\,beam$^{-1}$)&(mJy)&(mJy\,beam$^{-1}$)&(mJy)&$\alpha{^{1.3}_{4.5}}$\\
\hline
E1(2)&17$^{\rm s}\!\!$.913&46\farcs48&90.48&564.13&55.00&256.25&0.65\\
E2(5)&17.895&46.34&53.81&427.06&28.50&242.80&0.47\\
E3(6)&17.873&46.36&17.44&134.10&10.42&67.69&0.57\\
Core(7)&17.800&46.48&16.38&27.10&20.87&23.55&0.11\\
W1(8)&17.791&46.56&38.52&460.69&9.60&133.38&1.03\\
W2(11)&17.773&46.59&59.13&484.60&25.84&214.67&0.67\\
\hline
\end{tabular}
 \label{tab2}
\end{table*}
\end{center}
\section{Observations \& Image Processing}
\subsection{VLA plus Pie Town Observations}

Observations centred at 1359.518 MHz were taken with the VLA on December
15$^{\rm th}$ 2000 in A configuration using 64 channels across a bandwidth
of 3.125 MHz. Data from both right and left circular polarizations were
obtained.  The VLBA antenna at Pie Town also participated in the
observations and the IF signal was distributed to the VLA correlator via a
Western New Mexico Telephone Co. fibre optic link.  A single VLA antenna
was removed from the array so that its place in the electronics could be
used by the Pie Town antenna.  The length of the observing run was 8
hours. The strong calibrator 3C286 was employed for bandpass and absolute
flux calibration.

\subsection{MERLIN Observations} 

3C\,293 was observed with the UK MERLIN array (Thomasson 1986) with 7
telescopes on April 8$^{\rm th}$ 1998. The observations measured both left
and right hands of polarization across an 8\,MHz bandwidth centred upon
the redshifted frequency of the H{\sc i} line in 3C\,293 (1359 MHz). These
data were correlated into 64 frequency channels of width 125\,kHz
(28.8\,\kms). These observations were made over a period of 18\,hours and
were phase and bandpass calibrated using observations of the source OQ208
which were regularly interspersed throughout the observing run. An
observation of 3C\,286 at either end of the observing run was used to
calibrate the absolute flux density scale for all the MERLIN observations.
Initial calibration and editing of this data-set were made at Jodrell Bank
using local MERLIN software routines. Following this, these data were
imported into {\sc aips} where further editing and calibration values were
derived and applied using phase solutions obtained from the phase
calibration source OQ208 and using standard self-calibration techniques.
The observations and processing of the MERLIN data-set have been described
in more detail in Beswick et al. (2002) and Beswick (2002).

\subsection{VLBI Observations}

Global VLBI observations made with the European VLBI Network (EVN)
antennas at Westerbork, Onsala, Medicina, and Effelsberg, the VLBA, and
the phased VLA were obtained on November 18$^{\rm th}$ 1999.  The VLBA
correlator in Socorro produced 512 spectral channels across a bandwidth of
8\,MHz centred at 1359.29\,MHz.  Four level sampling was used and data
from both right and left circular polarizations were processed.  Delay,
rate and amplitude calibration were obtained from short (4 min)
observations of OQ208 obtained every $\sim$30\,min.  The length of the
observing run was 14 hours.  Bandpass calibration was obtained from
observations of the strong calibrator 3C345.

\subsection{Combination of these three data-sets}

Initial phase and amplitude calibration was done on each data-set
independently, using OQ208 in all cases, as described above. The MERLIN
data were then shifted to J2000 coordinates to match the other data-sets.  
The global VLBI data and VLA$+$PT data were averaged in frequency to
correspond with the lower velocity resolution MERLIN data, and the central
frequencies were shifted slightly so that the frequency range and channel
numbers in all data-sets matched exactly. The data-sets were then
concatenated by combining the VLA$+$PT and MERLIN data, and then the VLBI
data, with iterations of self-calibration at each stage.  The relative
weights for the data from each array were also checked during the process.  
The combined data were subsequently Fourier transformed and deconvolved
using a circular 30\,mas restoring beam to form a 2048$\times2048\times$23
spectral line cube to which standard spectral line routines within {\sc
aips} were applied.

The primary benefit of combining these data-sets is that it allows the
final image created from the radio interferometric data to sample a wide
range of angular size scales and thus give a more complete representation
of the radio emission.  The combined image with 30\,mas angular resolution
contains a total flux density of 3.2\,Jy compared with the total flux
density of 9.8\,Jy derived from the 200\,mas angular resolution MERLIN
image. The missing flux density in the higher angular resolution combined
image results from spatial filtering of the more diffuse structures which
are seen in the MERLIN image.

\begin{figure*}
\setlength{\unitlength}{1mm}
    \begin{picture}(80,105)
\put(0,0){\includegraphics{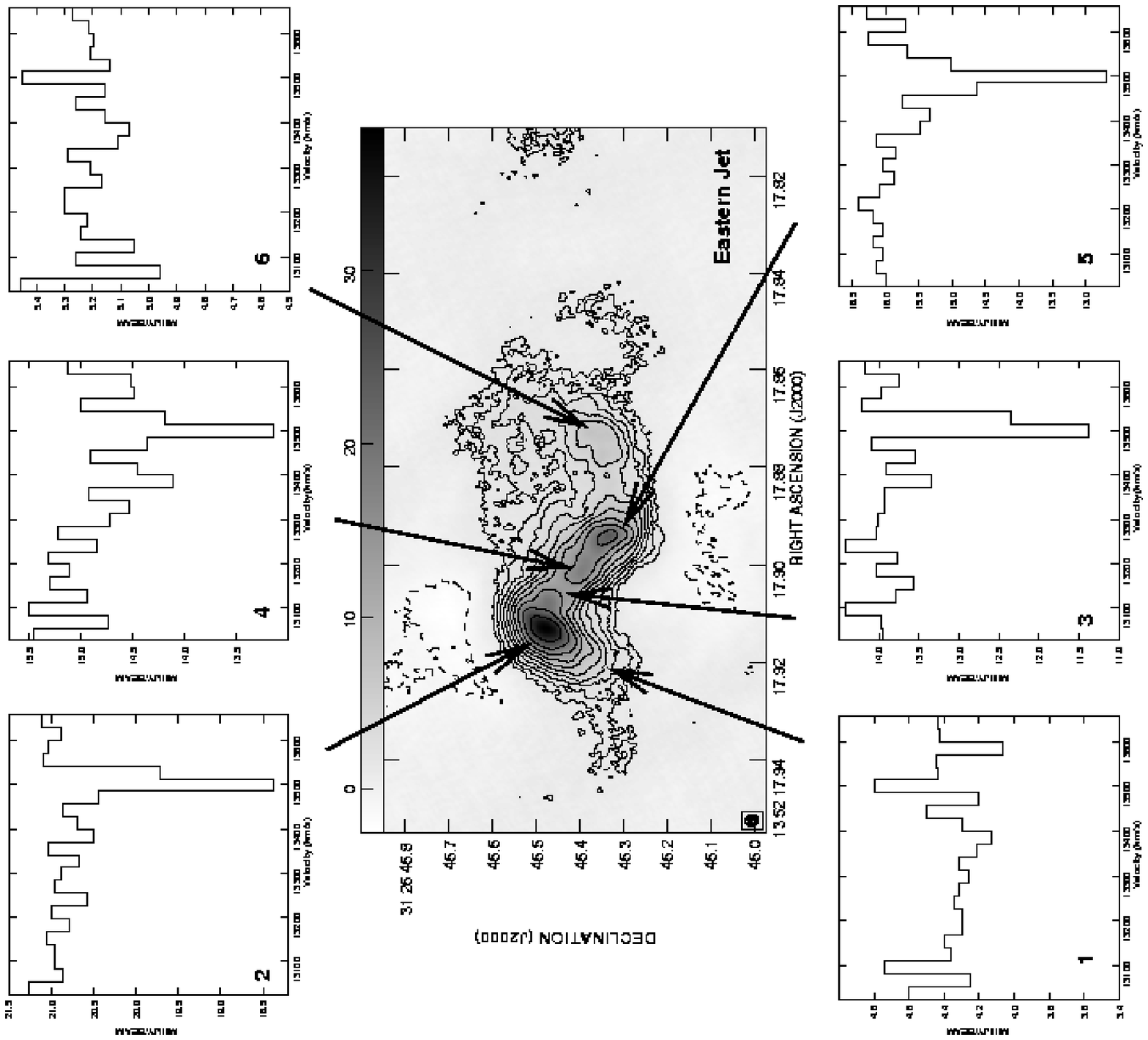}}
\end{picture} 
\vskip 14.5pc 
\caption{Montage of 1.35 GHz radio structure of the eastern part of the
source with a 30\,mas circular restoring beam along with selected H{\sc i}
absorption spectra. Contours are identical to Fig.\,3.}
\label{fig3}
\end{figure*}

\begin{figure*}
\setlength{\unitlength}{1mm}
    \begin{picture}(80,105)
\put(0,0){\includegraphics{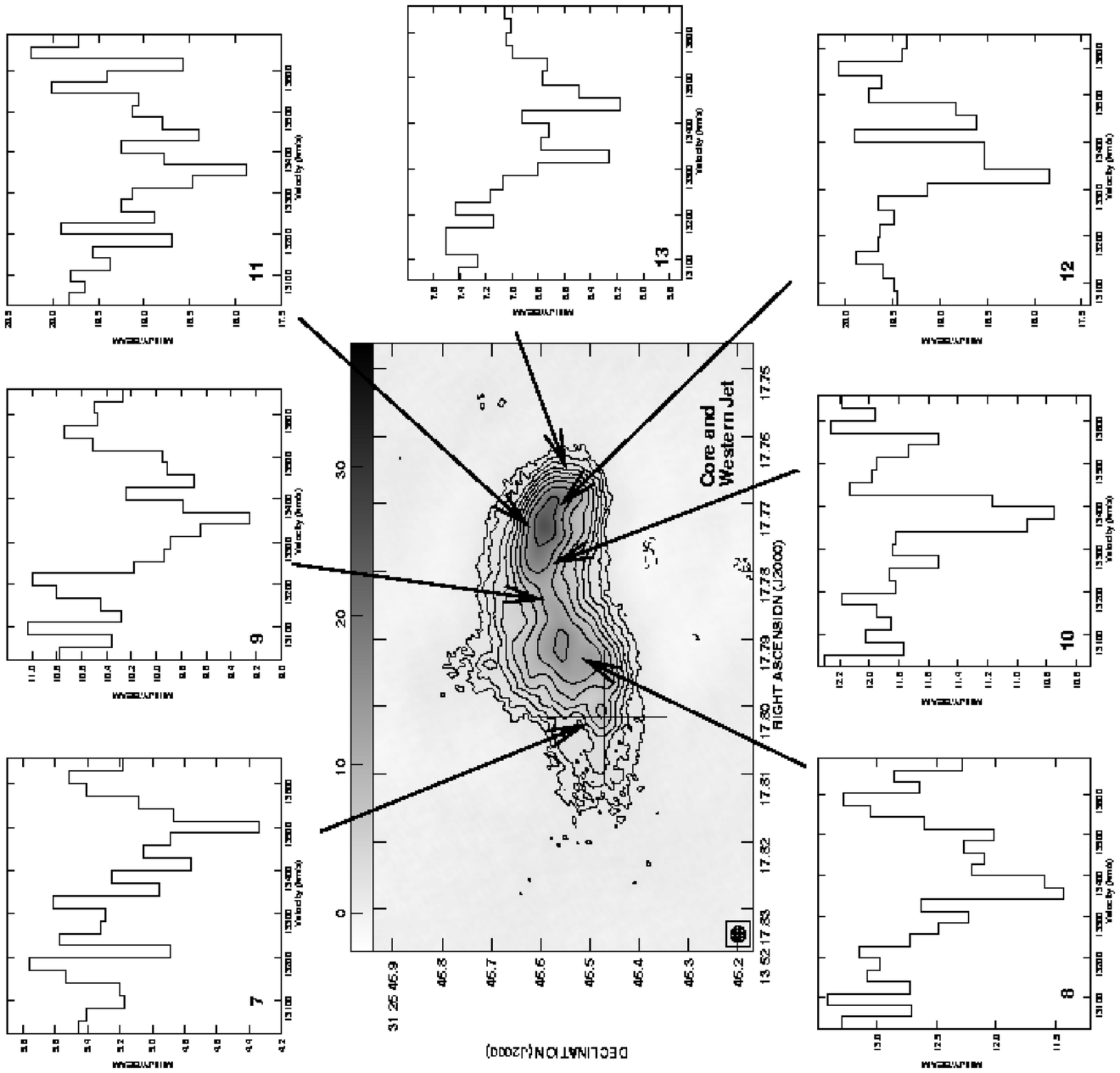}}
\end{picture} 
\vskip 14.5pc 
\caption{Montage of 1.35\,GHz radio emission toward the western part of the source with a
30\,mas circular restoring beam along with selected H{\sc i} absorption
spectra. Contours are identical to Fig.\,3. The cross shown marks the
position of the 5\,GHz core detected by Akujor et al. (1996). The size of the cross is much larger than the relative astrometric error between the 5\,GHz observations of Akujor et al. (1996) and the observations presented here.}
\label{fig4}
\end{figure*}

\section{Results}
\subsection{Multi-scale radio continuum emission}

The large scale radio continuum structure of 3C\,293 is shown in Fig.\,2.
This image was formed from VLA B-configuration data obtained as part of
the global VLBI observations and shows the large double jet/lobe structure
of the FR-II radio galaxy 3C\,293.  The radio continuum structure of
3C\,293 upon these angular scales consists of a wide two-sided radio
jet/lobe extending approximately 90\,arcsec at a P.A.$\sim$45$\degr$ with
a bright central radio core of peak flux density 3.44\,Jy. The radio
continuum structure of 3C\,293 upon these large scales shows the
north-western radio jet and lobe to have a significantly larger radio flux
than the south-eastern jet, consistent with the north-western jet
approaching upon these scales. The inset image in Fig.\,2 shows the
naturally weighted (0\farcs23$\times$0\farcs20 angular resolution) MERLIN
image of the central core region. This image, formed from the absorption
free continuum channels of the MERLIN data-set (see Beswick et al. 2002
for details of this observation), shows the structure of the central
3\,kpc of the radio jet of 3C\,293. This central region of 3C\,293 is
resolved into several continuum components following an east-west
orientation. The jet orientation between these two spatial scales deviates
by nearly 45$\degr$ as has been previously noted by Bridle et al. (1981),
Evans et al. (1999) and Beswick et al. (2002).
 
At the 0\farcs2 angular resolution of the MERLIN observations, the inner
jet of 3C\,293 shows at least four distinct radio components (labeled A,
B, C \& D in Fig.\,3 following Bridle et al. 1981) with two more
diffuse lobe-like components extending on either side of the central
region. At this resolution and frequency it is unclear which radio
continuum component is coincident with the AGN.  The lower portion of
Fig.\,3 shows the radio continuum structure of the inner part of the
MERLIN image. This continuum map, formed from the line-free portion of the
combined global VLBI, MERLIN and VLA spectral line cube, has been
convolved with a 30\,mas circular restoring beam and has a noise level of
0.4\,mJy\,beam$^{-1}$. In this high resolution image the inner kiloparsec
of the radio jet in 3C\,293 can be seen to deviate significantly from its
east-west trajectory.  The continuum components A, B, C \& D which are
barely resolved in the MERLIN image are composed of numerous compact,
bright knots of radio emission tracing the path of the radio jet. This jet
structure is consistent with the 50\,mas resolution 5\,GHz MERLIN results
presented by Akujor et al. (1996). The differences between the 5 GHz image
published by Akujor et al. (1996) and our higher resolution 1.3 GHz image
result from differences in the spectral index of the various components
that form the jet and the insensitivity to more diffuse structures of the
MERLIN 5\,GHz image presented by Akujor et al. due to sparseness of the
MERLIN array and exacerbated by the loss of the Defford telescope during their
observations. In particular, the weak 1.3\,GHz radio continuum component
at 13$^{\rm h}$52$^{\rm m}$17$^{\rm s}\!\!$.800,
31\degr26\arcmin46\farcs48 is at the same position as the steeply inverted
component identified by Akujor et al. as the radio core. Both these observations and those of Akujor et al. were phase referenced using the same nearby calibrator source providing good absolute positions which have been confirmed by fitting the positions of the bright inner hotspots (E1, E2 and W2) visible in both  data sets. Consequently the estimated positional offsets of these to observations is less than 15\,mas. The spectral
indices between 1.359 and 4.546\,GHz\footnotemark[1] and positions for
each of the components labelled in the lower part of Fig.\,3 are listed in
Tab.\,2, using S$_\nu\propto\nu^{-\alpha}$.

\begin{figure*}
\setlength{\unitlength}{1mm}
\begin{picture}(80,95)
\put(0,0){\includegraphics{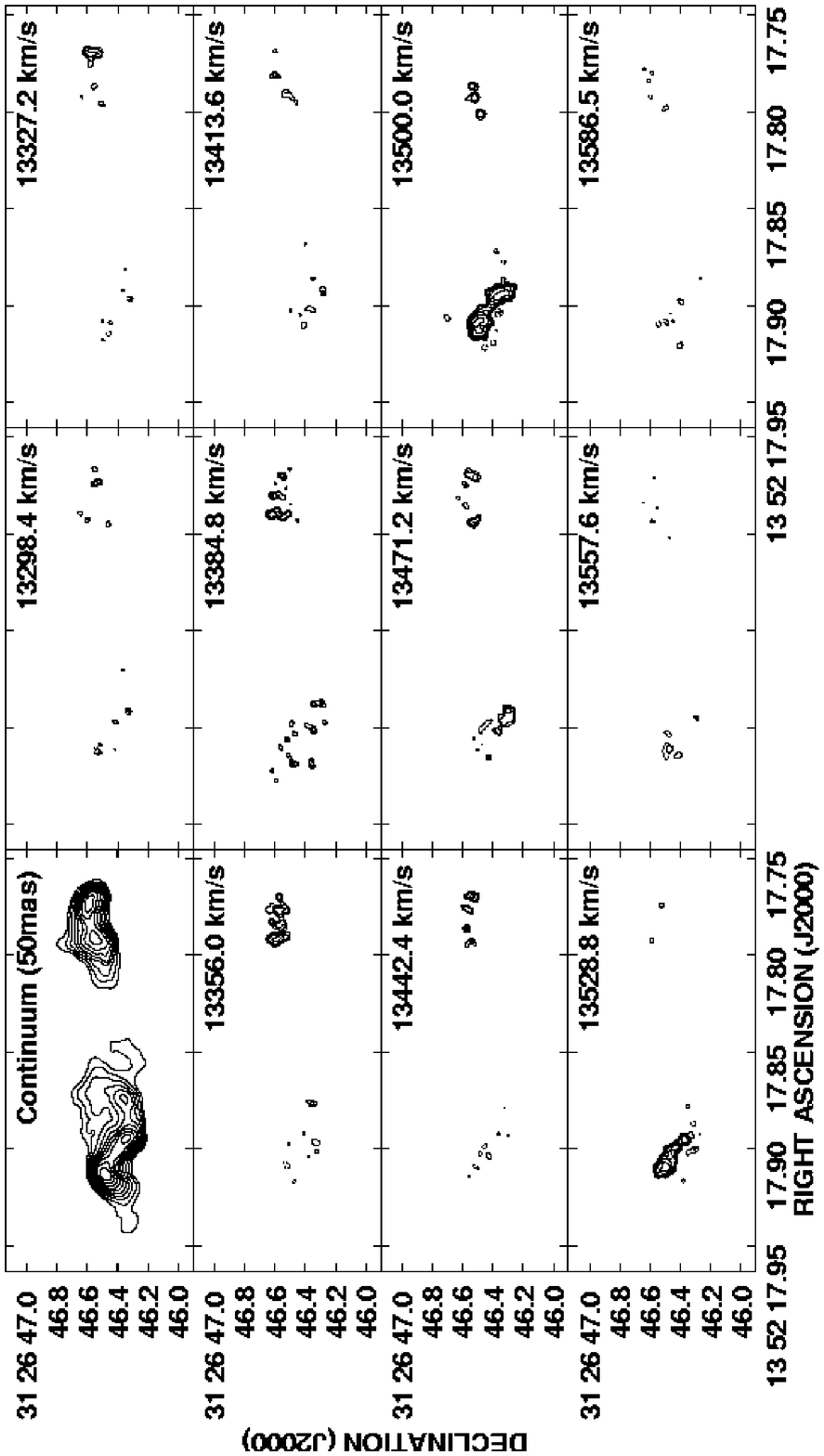}}
\end{picture}  
\caption{Nine individual channel maps of H{\sc i} absorption across the
inner kiloparsec of 3C\,293 with an angular resolution of 50 mas. The
contour levels mapping the absorption are $-$11.31, $-$8, $-$5.657, $-$4,
$-$2.828, $-$2, $-$1.414, $-$1, 1 and 1.414 times 2\,mJy\,beam$^{-1}$. The central
velocity of each channel is labeled. For spatial reference a continuum
map of the same area, also with an angular resolution of 50\,mas, is shown
in the top left hand panel. The contour levels on this continuum image are $\sqrt2$ times 3\,mJy\,bm$^{-1}$.}
\label{fig5}
\end{figure*}

\footnotetext[1] {MERLIN 4.5 GHz data with 50\,mas angular resolution has
been used for this comparison. Access to these data has been kindly
provided prior to publication by J. P. Leahy.}

\subsection{High resolution neutral hydrogen absorption observations}

Using the combined global VLBI, MERLIN and VLA data-set neutral hydrogen
absorption has been detected and resolved against the majority of the
radio continuum structure observed at mas angular resolutions. Figures 4
\& 5 show montages of selected H{\sc i} absorption spectra extracted from
the 30\,mas angular resolution data-set taken from the regions indicated
on the continuum images for both the eastern and western sections of the
inner jet structure shown in Fig.\,3. The noise levels on individual
channels of the 30\,mas resolution spectral line cube are
$\sim$0.7\,mJy\,beam$^{-1}$. The H{\sc i} absorption line characteristics
and the positions at which each of these 13 spectra occur are detailed in
Tab.\,3. Also in Tab.\,3 the peak opacities and associated H{\sc i} column
densities have been quoted. The column densities have been calculated
using N$_{\rm H}=$1.823$\times$10$^{20}\frac{\rm T_{spin}}{100\,{\rm K}}
\int \tau$\,dV and assuming a spin temperature of 100\,K. It should be
noted, however, that these values of N$_{\rm H}$ are likely to be lower
limits in some physical situations, such as close to the nuclei of active
galaxies or in outflows, where the gas spin temperature may be as large as
a few 1000\,K (Maloney, Hollenbach \& Tielens 1996).

As a result of the combination of three H{\sc i} absorption data-sets
(Global VLBI, MERLIN and VLA plus Pie Town) covering a wide range of {\it
u-v} spacings it is possible to image the H{\sc i} absorption reliably
over a variety of angular resolutions. Figure\,6 shows contour images of
nine individual velocity channels showing the position and velocity
distribution of the H{\sc i} absorption. In this figure the data-set has
been convolved with a 50\,mas circular beam in order to increase the
signal to noise of the absorption in the individual channels. From
Figs.\,4, 5, \& 6 it is clear that the velocity structure and line width
of the H{\sc i} absorption is significantly different against the the
eastern and western sections of the radio source. Against the western part
of the inner radio jet, the H{\sc i} absorption is significantly broader
and weaker than the absorption observed against the eastern components.
This is consistent with lower resolution H{\sc i} absorption observations
of this source by Haschick \& Baan (1985)  with the VLA and by Beswick et
al. (2002) with MERLIN.

The distribution of H{\sc i} optical depth and velocity toward the eastern
components of the inner radio jet are shown in Fig.\,7.  In Fig.\,7 the
optical depths of the narrow H{\sc i} absorption against the eastern
component are shown over the three spectral line channels that contain
absorption significantly greater than the noise levels in individual
channels. The spatial distribution of H{\sc i} in these three channels is
consistent with a strip of foreground gas at an approximate PA of
$\sim$35\degr. This strip of foreground gas is centred upon the rest
velocity of 3C\,293, 13\,500\,km\,s$^{-1}$, with a small south-west to
north-east velocity gradient visible over the three channels.

Against the western half of the inner jet structure the H{\sc i}
absorption is broader and generally composed of two or more velocity
components (see Fig.\,5). The complex velocity structure against this
component is presented in Fig.\,8 as three position velocity diagrams
plotted as a function of R.A.. These diagrams have been formed by
averaging the absorption signal convolved with three different restoring
beams (80, 50 and 30\,mas) over the declination range of the western radio
jet component.

\section{The radio continuum emission}
\subsection{The location of the AGN at 1.3\,GHz} 

In our previous sub-arcsecond MERLIN 1.3\,GHz images of 3C\,293 (Beswick
et al. 2002), it was not possible to distinguish the core component from
the extended radio continuum emission. However this source is well known
to have a steeply inverted core, which consequently is clearly visible as
an unresolved component in the 5 GHz MERLIN images presented by Akujor et
al. (1996). Akujor et al. also found this component to be unpolarized at
22 GHz providing further supporting evidence that it is the nuclear
component. The MERLIN 5 GHz position of this unresolved core identified as
the nucleus is marked by a cross on Fig.\,5 and listed in Tab.\,2. At 1.3
GHz, this component is only separable from the extended radio continuum
emission in our 30\,mas angular resolution image and has a low flux
density compared to the jet components within the image. Using Gaussian
fitting techniques, Akujor et al. determine that the core component is
unresolved with an angular size of $\ltsim$13\,mas. A similar analysis of
our 30\,mas angular resolution, 1.3\,GHz image also finds this component
is unresolved with a deconvolved size of $\ltsim$20\,mas implying a linear
size of $\ltsim$17\,pc.

The spectral index of the core between 1.359\,GHz and 4.546\,GHz at
50\,mas angular resolution is $\alpha^{1.3}_{4.5}=$0.11, using
S$_\nu\propto\nu^{-\alpha}$. This value is flat compared to that reported
by Akujor et al. between 15 and 22\,GHz ($\alpha^{15}_{22}=-$1.05). The
steeply inverted spectral index of $\alpha\approx-1$ can be inferred to
extend down to 5\,GHz when Akujor et al's high frequency observations are
compared with their 5\,GHz observations and new MERLIN 4.5\,GHz
observations.  The apparent flattening of the spectrum between 1.3 and
4.5\,GHz compared to higher frequencies, which is inferred from these
observations, is probably due to contributions to the derived 1.3\,GHz
flux density of this component by more diffuse non-nuclear radio emission
from the inner jet. These contributions will become more significant in
lower frequency observations since radio jets typically have a spectrum
which steepens as the synchrotron emitting particles in the jets age.  In
fact, if the true radio spectral index of the core component is assumed to
be $-$1 then from the 5\,GHz flux densities observed this would imply an
expected compact core flux density at 1.3\,GHz of $\sim$7\,mJy. In our
highest resolution (30\,mas) images presented here the fitted core flux
density is 13.5\,mJy, which implies either a break in the spectrum of the
radio core at theses frequencies or, more likely, that even with the
25\,pc linear resolution of these observations, that approximately half of
the recorded core flux density arises from contributions from unresolved
jet components.  Multi-frequency, VLBA continuum observations of the core of 3C\,293 have been proposed to isolate the core component from extended jet emission in the central tens of parsec.

It is also probable that the 1.3\,GHz flux density contribution of the
core component is further reduced by free-free absorption resulting from
intervening ionised gas expected to surround the AGN upon scales of
$<1$\,pc. In fact free-free absorption is observed in radio spectra of the
cores of many active galaxies, for example in many Seyfert nuclei ({\it
e.g.} Gallimore et al.  1999; Pedlar et al.  1998) and in similarly
gas-rich radio galaxies ({\it e.g.} 3C\,305; Jackson et al. 2003) at
frequencies $\ltsim$1.4\,GHz.  This implies that in these 1.3 GHz
observations the majority of the measured core flux density is contributed
by the jet emission rather than the true core.

\begin{center}
\begin{table*}
\caption[]{Summary of H{\sc i} absorption line properties. All velocities
are quoted in the optical heliocentric convention and have been obtained
from Gaussian fits to the spectral lines. Column densities have been
calculated using 1.823$\times$10$^{20}\frac{\rm T_{spin}}{100\,{\rm K}}
\int \tau$\,dV. In the following table and discussion throughout this
paper we assume a spin temperature of 100\,K. Limits upon the detected
H{\sc i} opacity detected in the 30\,mas data-set with a detection
threshold of 3$\sigma$ on the spectra averaged over an area of a few
beams.}
\begin{tabular}{l||c|c|c|c|c|c}
\hline
Component&R.A. (J2000)&Dec. (J2000)&Velocity\,(V$_{\rm c}$)&FWHM\,($\delta$V)&$\tau$&N$_{\rm H}$\\
&13$^{\rm h}$52$^{\rm m}$&31\degr\,26\arcmin&\kms&\kms&(peak)&$\times$10$^{20}$\,atoms\,cm$^{-2}$\\
\hline
1&17$^{\rm s}\!\!$.921&46\farcs35&-&-&$<$0.07&-\\
2(E1)&17.913&46.48&13497&52&0.132&6.9\\
3&17.905&46.44&13496&28&0.205&8.5\\
4&17.900&46.38&13485&33&0.124&5.2\\
5(E2)&17.895&46.34&13482&49&0.227&11.4\\
6(E3)&17.873&46.36&-&-&$<$0.06&-\\
7(Core)&17.800&46.48&13478&108&0.185&22.4\\
8(W1)&17.791&46.56&13365,\,13458&90,\,71&0.116&19.1\\
9&17.785&46.60&13338,\,13454&134,\,135&0.120&19.6\\
10&17.781&46.60&13356&61&0.105&10.0\\
11(W2)&17.773&46.59&13383,\,13496&169,\,85&0.096&16.2\\
12&17.767&46.56&13313,\,13441&54,\,28&0.091&11.7\\
13&17.763&46.53&13310,\,13441&54,\,28&0.139&29.0\\
\hline
\end{tabular}
 \label{tab3}
\end{table*}
\end{center}
\subsection{Inner jet structure and relative geometry of the radio jet}

The jet structure on sub-arcsecond angular resolution scales has been
discussed in detail by several authors ({\it e.g.} Bridle et al. 1981;
Akujor et al. 1996; Beswick et al. 2002). However the observations of the
inner kiloparsec region of the jet presented here provides the highest
angular resolution image of the inner jet region at any frequency.  The
spectral indices of the brightest jet components between 1.3 and 4.5\,GHz
are listed in Tab.\,2. With the exception of the core (discussed above),
these do not have inverted spectra and fall within the range of
0.5$\rightarrow$1. There are no significant spectral index gradients
detected along the path of the jet, however the western jet components
tend to have somewhat steeper spectral indices. At $\sim$GHz frequencies
the radio spectrum of a jet is affected by free-free absorption due to 
intervening ionised material along the line of sight to the observer and
by spectral ageing of electrons along the jet. Thus the slight steepening
of the spectral index against the western side of the jet can be
interpreted as an increase in free-free absorbing material in front of
this half of the source compared to the eastern side, superimposed upon
any spectral ageing effects. This interpretation is likely to be a
consequence of generally higher gas columns against the western side of
the source. This is consitent with the western jet receding as proposed by
Beswick et al. and Akujor et al. and supported by the detection of
one-sided optical jet emission in K-band {\it NICMOS} images (Leahy et al.
1999) which is coincident with the radio emission from components E1, E2
and E3.

Assuming that the sub-kpc scale jets are intrinsically symmetric and are
moving with relativistic velocities, the size of the counter-jet emission
(western side) is expected to appear smaller than that of the jet
emission. From the jet counter-jet arm ratio (see {\it e.g.} Giovannini et
al. 1998; Taylor and Vermeulen 1997) estimated from the image in Fig.\,3
(lower panel), we derive $\beta$cos$\theta$$\sim$0.55 where $\beta$ is
$v$/c, $v$ is the jet bulk velocity, and $\theta$ is the jet orientation with
respect to the line of sight.

The large gap between the core and E3 could be due to non-continuous
activity of the central AGN (but no gap is visible on the western side) or
more likely to a Doppler de-boosting effect. If the jet velocity is
$\sim$0.995c as found in other low and high power radio galaxies (see {\it
e.g.} Giovannini et al. 2001), the observed arm ratio implies
$\theta\sim$55$^\circ$ and the Doppler factor is =0.2. With this low
value, the observed jet brightness is too faint to be visible. In this
scenario the extended emission (E1, E2, E3, W1, and W2) is not the image
of the fast jet itself but low velocity shear layer emission surrounding
the fast jet spine. This interpretation is in agreement with the large
transversal size of the visible emission and with the brightness symmetry
of E and W components. The presence of a velocity structure in an
intermediate jet region (from the pc to the kpc scale) is expected by the
model proposed by Laing (1996) and related to the jet interaction with a
dense ISM (Giovannini 2003).

Upon larger scales, as shown in Fig.\,2, the $\sim$10 kiloparsec scale jet
of 3C\,293 has a P.A.$\sim$45\degr.  The apparently discordant
trajectories of the inner and outer jet structures imply a large position
angle change of the jet during this radio galaxy's lifetime. This position
angle change probably implies that the radio emission from 3C\,293 has
gone through two more outbursts and that the younger radio emission
observed from the inner jet might be a signature of the radio emission
from 3C\,293 being `born again' (Akujor et al 1996). In this case the
large scale jet emission would the remnant of older outbursts while the
bright inner jet emission is the result of an outburst of order 10$^5$
years old. The trigger for this latter outburst, possibly 3C\,293's
interaction with its nearby companion, may also have affected the
alignment of the radio emission from this latter outburst.

\section{Properties of the neutral hydrogen absorption} 

As has previously been noted by Beswick et al. (2002), the distribution of
H{\sc i} absorption against the central jet within 3C\,293 can be broadly
split into two differing gas components on the basis of the absorption
line widths and spatial distribution. As such, these two components can be
treated separately since they trace two different gas distributions and at
the resolution of these observations are spatially separated.

\subsection{Narrow H{\sc i} absorbing component: gas in front of the
eastern jet}

Against the eastern jet the H{\sc i} absorption lines have a narrow
velocity width ($\sim$40\,km\,s$^{-1}$ see Tab.\,3) centred at
$\sim$13500\,km\,s$^{-1}$ with no broad absorption lines detected against
this part of the source. Narrow H{\sc i} absorption lines are often
considered to be indicative of absorption that is the result of ambient
gas and hence are often attributed to gas lying at some distance from the
centre of the galaxy. In the case of 3C\,293 this is the most plausible
explanation for the location of the narrow H{\sc i} absorbing gas seen in
front of the eastern jet. In Fig.\,7 the areas of highest H{\sc i} opacity
are clearly shown to approximately follow a strip with a P.A. of
$\sim$35\degr. This distribution is, within astrometric errors, co-spatial
with the area of increased H{\sc i} optical depth observed in lower
resolution MERLIN H{\sc i} absorption observations, associated with the
location of foreground dust obscuration observed in {\it HST} images (see
figure\,4 \& section 4.1 of Beswick et al. 2002).

As can be seen from the absorption spectra (Fig.\,4) and the channel maps
presented in Fig.\,6, the velocity structure narrow absorption is only
spread over a few channels. However even with the velocity resolution
afforded by these observations ($\sim$28.8\,km\,s$^{-1}$\,channel$^{-1}$)
it can be seen in Fig.\,7 that the narrow absorption traces a small but
distinct velocity gradient of $\sim$50\,km\,s$^{-1}$\,arcsec$^{-1}$ in an
approximately south-west to north-east direction.  This velocity gradient
is also evident in the Gaussian fitted central velocities of the spectra
(Tab.\,3) and in Fig.\,6. This observed velocity gradient is only
traceable over an area of a few tenths of an arcsecond (limited by the
background continuum extent) but is consistent with the shallow H{\sc i}
absorption velocity gradient ($\sim$46\,km\,s$^{-1}$\,arcsec$^{-1}$)
observed over several arcseconds at lower resolution by Beswick et al.
(2002). Additionally, this shallow H{\sc i} structure is consistent with
the velocity gradients in ionised gas
($\sim$44\,km\,s$^{-1}$\,arcsec$^{-1}$ along a
PA$\sim$60$\rightarrow$65\degr) found using long-slit optical observations
of [O{\sc ii}] and [O{\sc iii}] emission lines by van Breugel et al.
(1984). Van Breugel et al's observations of the optical ionised gas traced
this $\sim$44\,km\,s$^{-1}$\,arcsec$^{-1}$ velocity gradient across the
optical extent of 3C\,293 out to radii of $\sim$10\arcsec, well beyond the
inner radio jets mapped in this experiment.

The co-spatial nature of the narrow H{\sc i} absorbing material against
the eastern jet with a foreground dust lane, along with its consistent
velocity structure compared to the optical emission line gas, implies that
all three of these components within the ISM of 3C\,293 are associated and
probably undergoing the same rotation.  Considering that the velocity
gradients observed in both the narrow H{\sc i} absorbing and the optical
emission line components are consistent and the physical extent over which
the optical velocity gradient is observed (out to a radius
$\sim$10\arcsec$\sim$8\,kpc), it is reasonable to conclude that both of
these components, along with the dust lanes, follow the galaxy's rotation.
Consequently it can be concluded that the dust, ionised gas and narrow
H{\sc i} absorption are situated on the nearside of the galaxy at a radius
of $\sim$8\,kpc and hence are not directly involved in the fuelling of the
central activity.

\begin{figure*}
\setlength{\unitlength}{1mm}
    \begin{picture}(80,93)
\put(0,0){\includegraphics{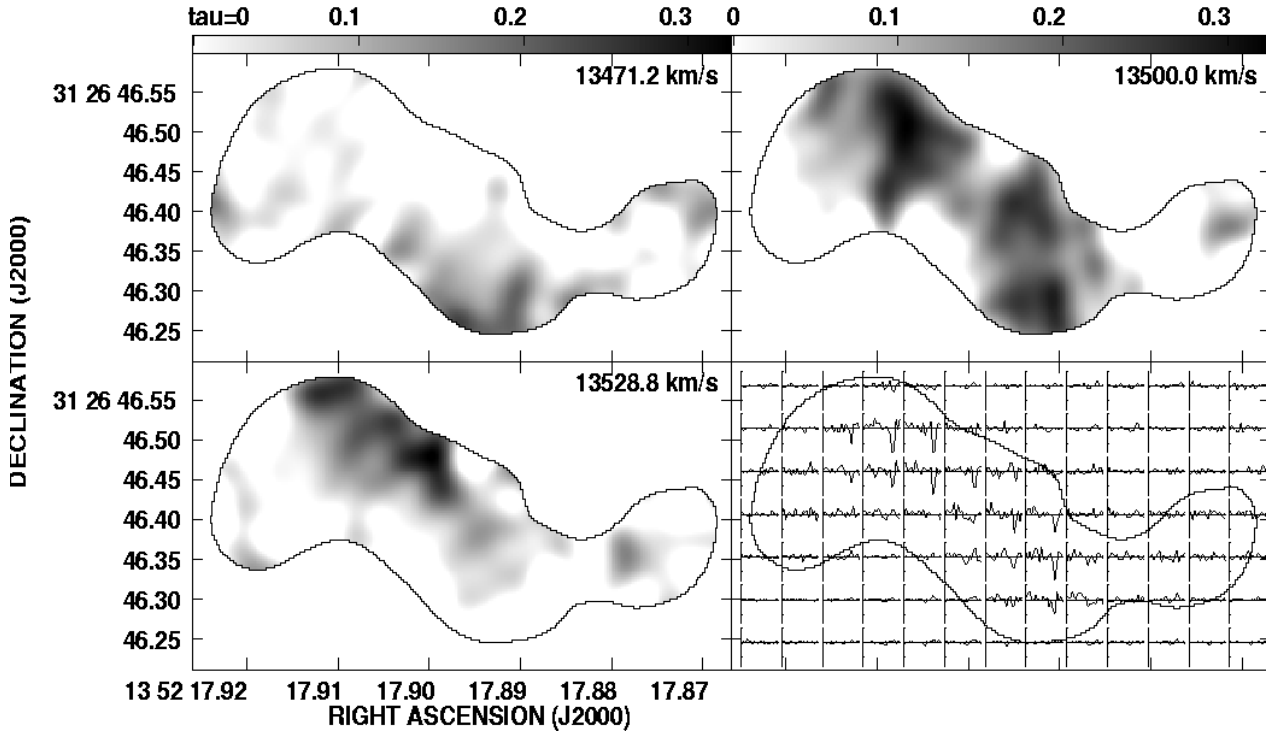}}
\end{picture}  
\caption{Optical depth toward the eastern jet for 3 channels over which
H{\sc i} absorption is detected at 50\,mas angular resolution. Bottom
right hand panel shows a grid of H{\sc i} spectra plotted against position
for the same area as the associated channel maps. Each spectrum is plotted
as flux density on a scale of -10 to 5\,mJy/beam against velocity over the
range 13200 to 13600\,km\,s$^{-1}$. Note that the ridge of higher H{\sc i}
opacity is aligned with a dust lane shown in {\it HST} observations (see
figure\,4 of Beswick et al. 2002, dust lane labeled no. 1).}
\label{fig6}
\end{figure*}

\subsection{Broad H{\sc i} absorbing component: gas in front of the
core and western jet}

Against the western portion of the VLBI scale jet structure, relatively
broad and complex H{\sc i} absorption is traced (Figs.\,5 \& 8). As we have
discussed in section\,4.2 the western jet is more deeply embedded within
the host galaxy than the eastern components and is viewed through higher
H{\sc i} gas column densities than the eastern jet.

The velocity structure of the H{\sc i} absorption against this portion of
the source is complex, as seen in the position velocity diagrams
(Fig.\,8).  It should initially be noted that the velocity structure
highlighted in this figure only represents a relativity small region of
gas in front of the radio continuum structure. However, Fig.\,8 does show
some structure within the absorbing gas, especially in the lower (80 and
50\,mas) angular resolution parts where both the absorbing gas and
illuminating continuum are less resolved. These velocity structures become
even more apparent in still lower resolution study of this source, such as
figure\,6 of Beswick et al. (2002). This H{\sc i} velocity distribution
can be interpreted in at least two ways. One interpretation of the data
presented is that the velocity structure against
the western half of the source is the
result of two gas systems centred at $\sim$13325 and 13500\,km\,s$^{-1}$
respectively. An alternative hypothesis is that the absorption traces a
velocity gradient in the gas in front of the AGN and sub-kiloparsec scale
jet, along with some gas clouds red or blueshifted away from this velocity
gradient.

\subsubsection{The complex, broad absorption as two separate gas components?} 

First, let us consider the hypothesis that the observed absorption against
the core and western part of the inner radio jet in 3C\,293 is composed of
two gas systems at different velocities. In this scenario the two gas structures are situated
along the line of sight to western half of the source at an undetermined
distance from nucleus. This interpretation provides some
explanation for the velocity structures observed in Fig.\,8 and, to a
lesser extent, those observed in figure\,6 of Beswick et al. (2002).  
The velocity structure seen in Fig.\,8 and in the Gaussian fits of spectra 8, 9, 10 and 11 in Table.\,3 can be satisfactorily fitted by two unassociated absorbing gas structures in the line of sight to the western jet and as such this must be the first explanation considered.

In section\,5.1 the narrow absorbing component against the eastern jet was discussed and interpreted to most likely be the result of ambient gas situated toward the nearside of the host galaxy. This component was shown to posses a small velocity gradient of $\sim$50\,km\,s$^{-1}$\,arcsec$^{-1}$. If it is assumed that this narrow component continues across the entire source, as is reasonable if this gas is associated with gas situated away from the centre of the source, this velocity gradient would imply that a narrow absorption component with a velocity of $\sim$13430, 13422 and 13410\,km\,s$^{-1}$ at positions core (7), W1(8) and W2(11) respectively should be observed. Although these velocity components are not directly matched by H{\sc i} absorbing components fitted to the spectra against the western jet (Table\,3) small variations in the gradient between the eastern and western halves of the source could be invoked in order to make the velocities consistent.  In this scenario the narrow absorption could account for one of the two composite gas structures that are observed against the western jet, with one other velocity component required to replicate the observed structure. This second velocity component can thus be accounted for by either gas localised toward the western jet or, more probably gas situated closer to the western jet and core structure in order that the projected angle of the inner jet (see section 4.2) results in the radio continuum components E1, E2 or E3 lying in front of the gas and thus not illuminating it.

\subsubsection{The complex, broad absorption as part of a rotating gas system?}

The broad and complex absorption against the western jet can also be interpreted as tracing gas in rotation about the nucleus of the galaxy. 

At resolutions of $\sim$1 arcsec, Haschick \& Baan (1985) only marginally
resolved the radio continuum in the core region of 3C\,293 into a
two-component structure. Haschick \& Baan identified the first of these
components with several high optical depth absorption components
($\tau>0.04$) which trace a velocity gradient of
83\,km\,s$^{-1}$\,arcsec$^{-1}$. This was interpreted as rotation of gas
within a disk or ring. With a factor of $\sim$7 higher resolution, Beswick
et al. (2002) also traced and began to resolve a still steeper gradient of
179\,km\,s$^{-1}$\,arcsec$^{-1}$ centred on the AGN. These conclusions are also supported by OVRO CO observations of a rotating asymetric ring of molecular gas surrounding the nuclear region (Evans et al. 1999). These lower resolution observations imply the existance of some form of rotating gas toward the centre of 3C\,293.   

If the results presented in this paper are also considered to trace gas rotating about the core as observed at lower resolution by Haschick \& Baan (1985), Evans et al. (1999) and Beswick et al. (2002), a similar, although a still steeper, velocity gradient might be expected to be observed. In this scenario the H{\sc i} absorption against the western jet and core should partially trace any gas rotating about the galaxy centre. If this is the case the best fit for these observations is a gradient of $\sim$410\,km\,s$^{-1}$\,arcsec$^{-1}$ centred upon the core and observed against a small area of the radio jet (see Fig.\,8).

Assuming that this steep velocity gradient is real, is it the inner part of the same rotating gas structure observed by both Haschick \& Baan's and Beswick et al. in H{\sc i} and Evans et al. in CO? Considering the angular resolution of both Haschick
\& Baan's and Beswick et al's previous experiments it is probable that, if
they were observing the same rotating ring of neutral gas, the angular
resolution of their experiments will have smeared the rotational
velocities observed. This effect will have resulted in them only observing a lower limit in true velocity gradient. If the gradient observed here is real these even higher angular
resolution H{\sc i} absorption observations trace a steeper velocity gradient, equivalent to 0.34\,km\,s$^{-1}$pc$^{-1}$. Unfortunately due to the lack of background
radio continuum adjacent to the eastern side of the core
this velocity gradient can only be traced over a few hundred parsecs against the core and the
western lobe.  Assuming the velocity gradient traced in Fig.\,8 is just a
section of the inner part of an inclined rotating disk of neutral gas and
that the disk is centred upon the core component, a dynamical lower limit
upon the enclosed mass within a radius of 400\,pc of $1.7\times10^9({\rm
sin}^{-2}i)\,{\rm cos}^{-1}\,\theta\,$M$_{\rm \odot}$ can be calculated,
where $i$ represents the inclination of the assumed disk and $\theta$ is
the angle between the major axis of the ring/disk and the axis over which
the position-velocity diagram has been average (east-west). This value
compares well with that derived over a larger radius by Beswick et al.
(2002).

The presence of a ring of neutral gas inferred from H{\sc i} absorption
observations has been reported in several previous cases in both radio
galaxies ({\it e.g.} NGC\,4261, van Langevelde et al. 2000 and 1946+708,
Peck, Taylor \& Conway 1999) and Seyfert galaxies (Gallimore et al. 1999)
ranging in size from a few tens of parsecs to a 100 or more parsecs.
Although the circumnuclear ring of H{\sc i} inferred in 3C\,293 is
considerably larger in extent than many other such examples, it has been
traced over a very wide range of scales from few tens of parsecs out to
greater than half a kiloparsec thus overlapping with the disk scales
previously observed in similar sources. This circumnuclear ring or disk of
neutral gas appears to be situated within the radius of an asymmetric molecular gas ring
detected in CO emission and absorption (Evans et al. 1999). Evans et
al. determined the radius of the molecular disk to be $\sim$2.8\,kpc and
containing $\sim$10$^{11}({\rm sin}^{-2}i)$\,M$_{\odot}$ of material
within this radius with $\sim$10$\%$ of this mass made up of molecular gas.
In these senses the neutral and molecular gas structures appear to mimic an `onion-skin' model in which a region of ionised gas surrounds the
AGN, and in turn is encompassed by a ring of neutral gas beyond which lies
the molecular gas. This model has been successfully used to explain the
gas structure of nuclear regions of Seyfert galaxies ({\it e.g.} Mundell
et al. 2003) albeit in lower power AGN and on much smaller scales.

\begin{figure}
\setlength{\unitlength}{1mm}
    \begin{picture}(80,100)
\put(0,0){\includegraphics{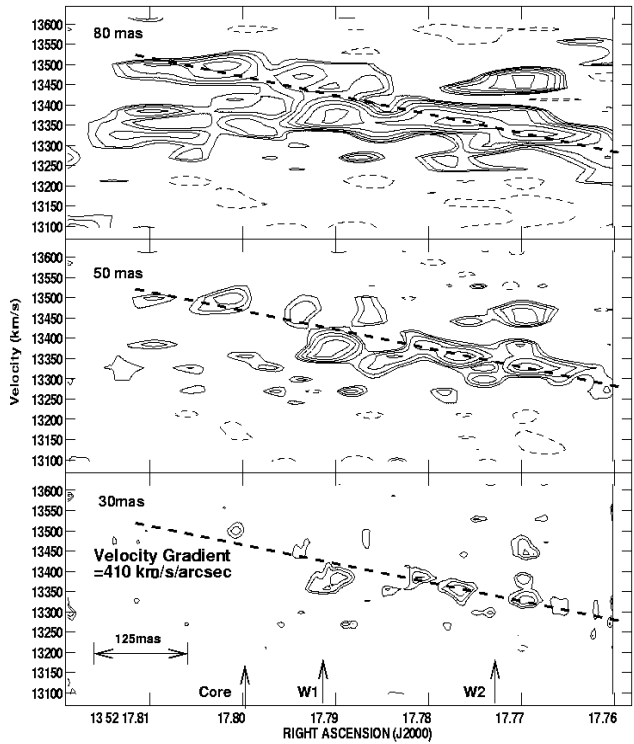}}
\end{picture}  
\caption{Multi-resolution position-velocity plots of H{\sc i} absorption
against the western jet component at the centre of 3C\,293. Contour levels
for all three plots are $-$16, $-$11.31, $-$8, $-$5.657, $-$4, $-$2.828,
$-$2, $-$1.414, $-$1 \& 1 times 0.4\,mJy\,beam$^{-1}$. In each of these
diagrams the absorption signal has been averaged over the declination
range of the continuum source. The spatial angular resolution of the
spectral line cubes that each of these position velocity plots have been
extracted from is labelled in the top left hand corner of each plot. The dashed line shown on all three plots represents a velocity gradient of 410\,km\,s$^{-1}$\,arcsec$^{-1}$. The
spatial position of radio continuum components labelled in Fig.\,3 are
also shown by arrows positioned along the bottom plot.}
\label{fig8}
\end{figure}

\subsection{Relationship of nuclear absorpion with gas inflows and outflows: interaction of the eastern jet with the ISM}

In the last few years one of the most interesting observational H{\sc i}
absorption results to emerge has been the detection of incredibly broad,
low optical depth, blue-wings in the absorption spectra of some active
galaxies ({\it e.g.} Oosterloo et al. 2000; Morganti et al. 2003).  In
particular, Morganti et al, using the WSRT observed a
$\sim$1000\,km\,s$^{-1}$ blue-shifted wing to the already broad H{\sc i}
absorption detected against the central ten kiloparsecs of 3C\,293. This
broad H{\sc i} absorption with typical opacities of $\sim$0.15$\%$
interpreted to be the result of outflows of gas from the central region,
which are probably driven by interactions between the radio plasma and the
ISM. Without the angular resolution to determine which part of the complex
radio jet structure (see Fig.\,3) illuminates and/or interacts with this
ISM component, Morganti et al. inferred from the
location of the broad optical emission lines (coincident with the eastern
jet components in Fig.\,3) that it is probable that broad H{\sc i} also
originates from in the vicinity of E1, E2 and E3 components.  As has
already been discussed, we surmise that this portion of the radio jet is
approaching, and it is also the location of a reported optical jet feature
(Leahy et al. 1999). The trajectory of this portion of the jet, along with
its location buried deep within a dust lane in the central few kiloparsecs
of the host galaxy do support their hypothesis. Additionally as we describe in section 4.1.2 the radio jet emission observed is consistent not with a fast jet but with emission from a low velocity shear layer surrounding the faster jet spine. This may provide an environment in which a jet-ISM interaction can occur and accelerate material to the high velocities seen by Morganti et al. whilst also allowing regions to cool and recombine to become neutral while moving at high velocities. This part of the inner jet in 3C\,293 has previously been suggested as the site for probably jet-ISM interactions and the entrainment of gas by van Breugel et al. (1984). Unfortunately neither the observations
presented here or those presented in Beswick et al. (2002) have the
sensitivity or sufficient bandwidth to confirm the broad
H{\sc i} absorption detected with WSRT and hence cannot be used to
categorically pinpoint the spatial location of this blue-shifted gas.
However it is clear that this broad absorption component is probably not
related to the relatively narrow and deep nuclear absorption detected in
this and by previous H{\sc i} observations of this source ({\it e.g.}
Haschick \& Baan 1985; Beswick et al. 2002).

\section{Conclusions}

We have used observations made using the VLA including Pie Town, MERLIN
and global VLBI to study the 1.3\,GHz radio continuum structure of the
central kiloparsec of the peculiar radio galaxy 3C\,293 and to investigate
the kinematics and the distribution of H{\sc i} via absorption against
this radio continuum.

 We confirm the component identified originally by Akujor et al. (1996) as
the most probable site of the central engine and place limits upon its
size of $\ltsim$17\,pc. Using both MERLIN 4.5\,GHz data (Akujor et al.
1996) and new high resolution 1.3\,GHz observations presented here we
determine a radio spectral index of $\alpha^{1.3}_{4.5}=$0.11 for the core
component. We discuss in detail the differences in the value of this
spectral index compared to that derived for the same component at higher
radio frequencies ($\alpha^{15}_{22}\approx-1$; Akujor et al. 1996). It is
concluded the flat spectral index derived in this study may be affected by
a significant amount of unresolved radio emission from the jet close to
the AGN and that this has been incorporated in our 1.3\,GHz flux density
measurement of the component resulting in the observed flat spectrum. In
addition to the core we have mapped the radio continuum emission of the
jet in 3C\,293 across a variety of angular scales. From this multi-scale
approach it is apparent that the trajectory of the radio jet emission in
3C\,293 has changed significantly over the source's lifetime.  In this
region, observational data suggest the presence of an intrinsically
symmetric jet with a highly relativistic spine, surrounded by a low
velocity shear layer because of the jet interaction with the dense ISM.  
The jet orientation with respect to the line of sight is $\sim$50$^\circ$
with the eastern jet approaching us.

Extensive H{\sc i} absorption has been detected against both the eastern
and western jet components within the central kiloparsec of 3C\,293,
consistent with lower angular resolution studies by Haschick \& Baan
(1985) and Beswick et al. (2002). As was previously known, the structure of the H{\sc i} absorption against the
eastern jet components primarily consists of strong and narrow features
with a small velocity gradient whereas the absorption against the western
and core components are much broader.

The narrow H{\sc i} absorption detected against the eastern radio jet
traces a small velocity gradient of $\sim50$\,km\,s$^{-1}$\,arcsec$^{-1}$,
consistent with the velocity gradient observed in ionised gas by van
Breugel et al. (1984). Additionally we have re-confirmed (following
Beswick et al. 2002) that this narrow H{\sc i} absorption is co-spatial
with the location of dust lanes observed by the {\it HST}. From the
association of these three components we conclude that they are all
most likely situated $\sim$8\,kpc from the central part of the galaxy and
are all probably undergoing galactic rotation.

Against the western radio jet and core complex H{\sc i} absorption is also
detected. This absorption is discussed in terms of either tracing two gas structures at undetermined distances along the line of sight to the jet or a steep velocity gradient which may be interpreted as neutral gas in rotation about the core.  If this is interpreted as rotation by a gas disk, it would imply an enclosed mass of  at least 1.7$\times$10$^9$  solar masses within a radius of 400\,pc of the core.

\section*{Acknowledgments}

RJB acknowledges PPARC support. ABP thanks the staff of Jodrell Bank
Observatory for their hospitality, and acknowledges support from MPIfR
during this project.  We would also like to thank J. P. Leahy for useful
discussions about this project and allowing us access to his data prior to
publication. We thank A. Pedlar for his numerous contributions to the
earlier stages of this project and encouragement throughout. The authors
also thank the referee for many useful comments that
have helped to improve the content and structure of this publication.

The VLA and VLBA are operated by the National Radio Astronomy Observatory.
The National Radio Astronomy Observatory is a facility of the National
Science Foundation operated under cooperative agreement by Associated
Universities, Inc. MERLIN is a national facility operated by the
University of Manchester on behalf of PPARC in the UK.  The European VLBI
Network is a joint facility of European, Chinese, South African and other
radio astronomy institutes funded by their national research councils.
 
\bibliographystyle{mnras}

\begin{thebibliography}{}
\bibitem[\protect\citename{Akujor et al.,  }%
        1996]{akujor96}  Akujor C. E., Leahy J. P., Garrington S. T., Sanghera H., Spencer R. E., Schilizzi R. T., 1996, MNRAS, 278, 1
\bibitem[\protect\citename{Allen et al.,  }%
        2002]{allen02}  Allen M. G., Sparks W. B., Koekemoer A., Martel A. R., O'Dea C. P., Baum S. A., Chiaberge M., Macchetto F. D., Miley G. K. 2002 ApJS, 139, 411
\bibitem[\protect\citename{Baan \& Haschick,  }%
        1981]{baan81}  Baan W. A., Haschick A. D., 1981, ApJ, 243, L143
\bibitem[\protect\citename{Beswick et al.,  }%
        2002]{beswick02}  Beswick R. J., Pedlar, A., Holloway, A. J.,  2002, MNRAS, 329, 620
\bibitem[\protect\citename{Beswick,  }%
        2002]{beswick02thesis}  Beswick R. J. 2002, PhD thesis, Univ. Manchester, UK.
\bibitem[\protect\citename{Bridle et al.,  }%
        1981]{bridle81}  Bridle A. H., Fomalont E. B., Cornwell T. J., 1981, AJ, 86, 1294
\bibitem[\protect\citename{Cecil et al., }%
        2001]{cecil01} Cecil G., Bland-Hawthorn J., Veilleux S., Filippenko, A. V., 2001, ApJ, 555, 338
\bibitem[\protect\citename{Evans et al., }%
        1999]{evans99}  Evans A. S., Sanders D. B., Surace J. A., Mazzarella J. M., 1999, ApJ, 511, 730
\bibitem[\protect\citename{Fanaroff \& Riley., }%
        1974]{fanaroff74}  Fanaroff B. L., Riley J. M., 1974, MNRAS, 167, 31
\bibitem[\protect\citename{Gallimore et al., }%
        1999]{gallimore99}  Gallimore J. F., Baum S. A., O'Dea C. P., Pedlar A., Brinks E., 1999, ApJ, 524, 684
\bibitem[\protect\citename{Giovannini et al., }%
        1998]{giovannini98}  Giovannini G., Cotton W. D., Feretti L., Lara L.,  Venturi T., 1998, ApJ, 493, 632
\bibitem[\protect\citename{Giovannini et al., }%
        2001]{giovannini01}  Giovannini G., Cotton W. D., Feretti L., Lara L.,  Venturi T., 2001, ApJ, 552, 508
\bibitem[\protect\citename{Giovannini, }
        2003]{giovannini03}  Giovannini G., 2003, in {\it ``The Physics of Relativistic Jets in the Chandra and XMM Era",} eds G. Brunetti, D.E. Harris, R.M. Sambruna, G. Setti; New Astronomy Reviews, Vol. 47, p551
\bibitem[\protect\citename{Haschick \& Baan,  }%
        1985]{haschick85}  Haschick A. D., Baan W. A., 1985, ApJ, 289, 574
\bibitem[\protect\citename{ Heckman et al., }%
        1985]{heckman85}  Heckman T. M., Illingworth G. D., Miley G. K.,  van Breugel W. J. M., 1985, ApJ, 299, 41
\bibitem[\protect\citename{Jackson et al.,  }%
         2003]{Jackson03} Jackson N. J., Beswick R. J., Pedlar A.,  Cole G. H. J., Sparks W. B., Leahy J. P., Axon D. J., Holloway A. J., 2003, MNRAS, 338, 643
\bibitem[\protect\citename{Laing, }%
          1996]{laing96} Laing R. S., 1996, in {\it ``Energy Transport in Radio Galaxies and Quasars'',} eds. Hardee P. E., Bridle A. H., Zensus J. A., Astron. Soc. Pac., Vol. 100, San Francisco, p. 241
\bibitem[\protect\citename{Leahy et al., }%
          1999]{leahy99} Leahy J. P., Sparks W. B., Jackson N. J. F., 1999, AAS, 194, 7304L
\bibitem[\protect\citename{Maloney et al.,   }%
        1996]{maloney96} Maloney P. R., Hollenbach D. J., Tielens A. G. G. M1996, ApJ, 466, 561
\bibitem[\protect\citename{Martel et al.,   }%
        1999]{martel99}  Martel A. R., et al. 1999, ApJS, 122, 81
\bibitem[\protect\citename{Morganti et al.,   }%
        2003]{morganti03}  Morganti R., Oosterloo T. A., Emonts B. H. C., van der Hulst J. M., Tadhunter C. N., 2003, ApJ, 593, L69
\bibitem[\protect\citename{Mundell et al.,   }%
        2003]{mundell03}  Mundell C. G., Wrobel J. M., Pedlar A., Gallimore J. F.  2003, ApJ, 583, 192
\bibitem[\protect\citename{Oosterloo et al.,   }%
        2000]{osterloo00}  Oosterloo T. A., Morganti R., Tzioumis A., Reynolds J., King E., McCulloch P. \& Tsvetanov Z. 2000, AJ, 119, 2085
\bibitem[\protect\citename{Peck et al.,   }%
        1999]{peck99}  Peck A. B., Taylor G. B., Conway J. E. 1999, ApJ, 521, 103
\bibitem[\protect\citename{Pedlar et al.,   }%
        1998]{pedlar98}  Pedlar A., Fernandez B., Hamilton N. G., Redman M. P., Dewdney P. E. 1998, MNRAS, 300, 1071
\bibitem[\protect\citename{Shostak et al., }%
                     1983]{shostak83} Shostak G. S., van Gorkom J. H. Ekers R. D., Sanders R. H., Goss W. M., Cornwell T. J., 1983, A \& A, 119, L3
\bibitem[\protect\citename{Taylor \& Vermeulen,  }%
         1997]{taylor97} Taylor G. B., Vermeulen R. C. 1997, ApJ,485, L9
\bibitem[\protect\citename{Thomasson,  }%
         1986]{thomasson86} Thomasson P., 1986,QJRAS, 27, 413
\bibitem[\protect\citename{van Breugel et al.,  }%
         1984]{vanbreugel84} van Breugel W., Heckman T., Butcher H., Miley G., 1984, ApJ, 277, 82
\bibitem[\protect\citename{van Langevelde et al.,  }%
         2000]{vanbreugel2000} van Langevelde H. J., Pihlstr\"{o}m Y. M., Conway J. E., Jaffe W., Schilizzi R. T., 2000, A \& A 354, L45
\label{lastpage}
\end{thebibliography}

\end{document}